\documentclass[twocolumn,secnumarabic,amssymb,nobibnotes,aps,prl,superscriptaddress]{revtex4-1}

\usepackage[colorlinks=true,urlcolor=blue]{hyperref}
\usepackage{letltxmacro}
\usepackage{longtable}
\usepackage{amsfonts}
\usepackage{graphicx}
\usepackage{hyperref}
\usepackage{epstopdf}
\usepackage{wrapfig}
\usepackage{amsmath}
\usepackage{amssymb}
\usepackage{dcolumn}
\usepackage{float}
\usepackage{color}
\usepackage{bm}

\definecolor{red}{rgb}{0.75,0,0}
\definecolor{blue}{rgb}{0,0,0.75}
\definecolor{green}{rgb}{0,0.5,0}

\graphicspath{{Figures/}}

\hypersetup{colorlinks,bookmarksopen,bookmarksnumbered,
citecolor=black,
linkcolor=black,
pdfstartview=green,
urlcolor=black}

\newcommand{\D}{{\rm d}}

\newcommand{\Hthree}{\mathbb{H}^3}
\newcommand{\Sthree}{\mathbb{S}^3}
\newcommand{\Rthree}{\mathbb{R}^3}
\newcommand{\Htwo}{\mathbb{H}^2}

\newcommand{\Rpa}{R}
\newcommand{\dpa}{d}
\newcommand{\dmax}{d^{*}}
\newcommand{\Ebe}{E_{\mathrm{be}}}
\newcommand{\Ead}{E_{\mathrm{ad}}}
\newcommand{\Eto}{E_{\mathrm{to}}}
\newcommand{\Aad}{A_{\mathrm{ad}}}

\newcommand{\Rw}{R_{w}}

\newcommand{\Eunwrap}{E_{\mathrm{U}}}
\newcommand{\Epartial}{E_{\mathrm{P}}}
\newcommand{\Efull}{E_{\mathrm{F}}}

\newcommand{\ktheta}{C_{\perp}}

\LetLtxMacro{\originaleqref}{\eqref}
\renewcommand{\eqref}{Eq.~\originaleqref}

\def\be{\begin{equation}}
\def\ee{\end{equation}}

\def\bmul{\begin{multline}}
\def\emul{\end{multline}}

\def\bea{\begin{eqnarray}}
\def\eea{\end{eqnarray}}

\begin{document} 

\title{Metastability of lipid necks via geometric triality}

\author{Piermarco Fonda}
\email{piermarco.fonda@mpikg.mpg.de}
\affiliation{Theory \& Bio-Systems, Max Planck Institute of Colloids and Interfaces, Am M\"uhlenberg 1, 14476 Potsdam}
\affiliation{Instituut-Lorentz, Universiteit Leiden, P.O. Box 9506, 2300 RA Leiden, Netherlands}
\author{Luca Giomi}
\email{giomi@lorentz.leidenuniv.nl}
\affiliation{Instituut-Lorentz, Universiteit Leiden, P.O. Box 9506, 2300 RA Leiden, Netherlands}

\begin{abstract}
``Necks'' are features of lipid membranes characterized by an uniquley large curvature, functioning as bridges between different compartments. These features are ubiquitous in the life-cycle of the cell and instrumental in processes such as division \cite{Storck2018}, extracellular vesicles uptake \cite{Mulcahy2014} and cargo transport between organelles \cite{Bonifacino2004}, but also in life-threatening conditions, as in the endocytosis of viruses \cite{Votteler2013,Rossman2013} and phages \cite{Rajaure2015}. Yet, the very existence of lipid necks challenges our understanding of membranes biophysics: their curvature, often orders of magnitude larger than elsewhere \cite{Avinoam2015}, is energetically prohibitive, even with the arsenal of molecular machineries and signalling pathways that cells have at their disposal. Using a geometric {\em triality}, namely a correspondence between three different classes of geometric objects, here we demonstrate that lipid necks are in fact metastable, thus can exist for finite, but potentially long times even in the absence of stabilizing mechanisms. This framework allows us to explicitly calculate the forces a corpuscle must overcome in order to penetrate cellular membranes, thus paving the way for a predictive theory of endo/exo-cytic processes. 
\end{abstract}

\maketitle

Because of their fundamental role in cellular remodelling, lipid necks have drawn considerable interest in biophysics over the past three decades. Experimentally, great effort has been made toward reproducing and manipulating necks in artificial lipid bilayers, with the goal of reducing the seemingly endless complexity of cellular matter to a small number of control parameters (e.g. Refs. \cite{Haluska2006,VandenBogaart2010,VanderWel2016,Steinkuhler2020}). On the theoretical front, much attention has been devoted to understanding how necks can be formed and stabilized. A major outcome of this endeavor is the so-called {\em neck stability condition}, relating the neck curvature and the spontaneous curvature of the lipid bilayer \cite{Seifert1991a,Fourcade1994} (see also Refs. \cite{Julicher1996,Agudo-Canalejo2016} for generalizations). Other studies, have instead identified in the adhesion with a solid substrate a possibile stabilizing mechanism for necks formation \cite{Deserno2004,Zhang2015}. Whether ascribed to spontaneous curvature or external forces, these results agree on the conclusion that lipid necks can only exist in the presence of stabilizing mechanisms, which act to counterbalance the enormous bending moments resulting from necks' nanometer-sized radius of curvature. 

In this article, we address the problem at its fundations and demonstrate that lipid necks are in fact {\em metastable} and can exist for a finite, but potentially long, time, even in the absence of stabilizing mechanisms. Our approach is rooted in a geometric {\em triality}, namely a correspondence between three classes of geometric objects (i.e. Willmore surfaces, minimal surfaces and Euler {\em Elastica}) embedded in as many topological spaces (i.e. $\Rthree$, $\Hthree$ and $\Htwo$), and capitalizes on tools and concepts from apparently disconnected areas of mathematics and high energy physics, such as the geometry of curves embedded in space forms \cite{Langer1984a} and holographic entanglement in the context of the AdS/CFT correspondence of string theory \cite{Ryu2006}. 

\begin{figure}[t]
\includegraphics[width=\columnwidth]{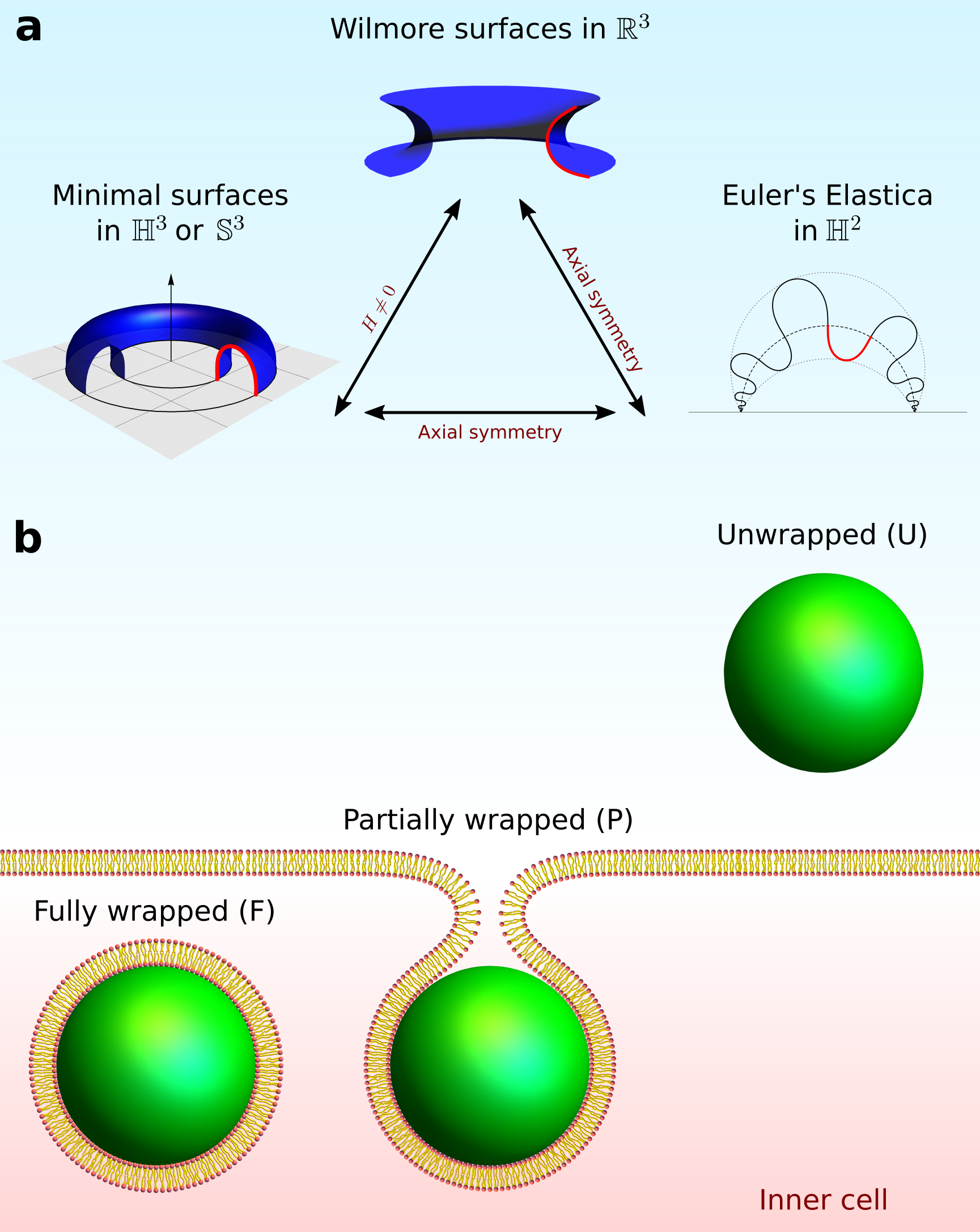}
\caption{
\textbf{a)} Diagram illustrating the geometric triality expressed by Eqs. (\ref{eq:E_be}), (\ref{eq:weyl bending}) and (\ref{eq:elastica bending}). \textit{Top}: Axisymmetric Willmore surface connecting a sphere and a plane. \textit{Left}: Axisymmetric hyperbolic minimal surface in the Poicar\'e half-space representation of $\Hthree$. The surface ends on two concentric circles at the conformal boundary, shown as a gray flat plane \cite{Fonda2015a}. \textit{Right}: wavelike elastica in the Poincar\'e half-plane oscillating between two constant geodesic curvature curves (dotted lines) and centred on a $\Htwo$ geodesic (dashed line) \cite{Fonda2017}. The horizontal straight line is the conformal boundary. The red lines in the three insets highlight the same geometrical object. \textbf{a)} The three possible states for a nanoparticle in the proximity of a membrane: unwrapped (U), partially wrapped (P) and fully wrapped (F).} 
\label{fig:one}
\end{figure}

The central quantity in the mechanics of lipid membranes is the bending energy
\be
\Ebe= 2 \kappa \int_S \D A\,H^2 \;, 
\label{eq:E_be}
\ee
where $\kappa$ is the bending modulus, $H=(C_1+C_2)/2$ is the local mean curvature, $C_i$ are the principal curvatures and the integral spans the mid-surface $S$ of the lipid bilayer \cite{Canham1970,Helfrich1973}. \eqref{eq:E_be} can be augmented with further terms to enforce global constraints (e.g. fixed area and volume), or with additional energy costs, such as those resulting from surface tension, spontaneous curvature, adhesion etc. Unlike $\Ebe$, however, these contributions scale with the system size and become irrelevant at the length scale of necks. By contrast, the bending energy, \eqref{eq:E_be}, is invariant with respect to scale transformations of the form $\bm{r}\rightarrow\lambda\bm{r}$, with $\bm{r}$ the surface position and $\lambda>0$ a scaling factor, and inevitably becomes dominant at sufficiently small scales. The problem of neck stability, thus reduces to finding minimizers of $\Ebe$, also known as Willmore surfaces \cite{Willmore1965}, with cylindrical topology. The latter, however, is a formidable task, owing to the fact that minimizing $\Ebe$ yields a nonlinear fourth-order partial differential equation in $\bm{r}$ \cite{Guckenberger2017}. 

These difficulties can be overcome by taking advantage of another fundamental symmetry of \eqref{eq:E_be}, namely {\em conformal invariance} \cite{SI}. The first striking consequence of this symmetry is that $\Ebe$ transforms as follows under special rescalings of the ambient space metric known as Weyl transformations \cite{Alexakis2010,Fonda2015}:
\be
\Ebe =  \tilde{E}_{\mathrm{be}} \pm \frac{2\kappa}{L^{2}}\,\tilde{A} \;,
\label{eq:weyl bending}
\ee
where $\tilde{E}_{\mathrm{be}}$ and $\tilde{A}$ are respectively the bending energy and the area of $S$ when embedded in either the three-sphere $\mathbb{S}^3$ (plus sign) or in the three-dimensional hyperbolic space $\Hthree$ (minus sign). $L$ is the characteristic length scale of these spaces, defined from the Ricci scalar $\mathcal{R}=\pm 6/L^2$. From this it follows that area-minimizing (i.e. {\em minimal}) surfaces in either $\Sthree$ or $\Hthree$ are also extrema of the bending energy \eqref{eq:E_be}.

Furthermore, in the special case of surfaces of revolution, obtained upon revolving a plane curve $\gamma$ around a fixed axis, $\Ebe$ can be expressed in terms of the bending energy of $\gamma$, when this is embedded in the two-dimensional hyperbolic plane $\Htwo$ \cite{Bryant1986}:
\be
\Ebe =  \pi\kappa \left( 4\chi + \ell \int_\gamma \D \tau\,k^{2}\right)\;,
\label{eq:elastica bending}
\ee
where $\chi$ is the Euler characteristic of $S$, $\ell$ the length scale of $\Htwo$, with $\mathcal{R}=-2/\ell^2$ the Ricci scalar, $\D\tau=\D s\,\ell/r$ is the hyperbolic line element, with $s$ the Euclidean arc-length, and $k$ the geodesic curvature of $\gamma$ when embedded in $\Htwo$. From \eqref{eq:elastica bending} it follows that axisymmetric minimizers of $\Ebe$ are {\em Elastica} curves in $\Htwo$ \cite{Langer1984a}. Together, Eqs. (\ref{eq:weyl bending}) and \eqref{eq:elastica bending} provide a dictionary that allows us to describe the same object, namely the mid-surface of a lipid neck, in three different languages: the geometry of Willmore surfaces in $\Rthree$, of minimal surfaces in $\Hthree$ or $\Sthree$ and of {\em Elastica} curves in $\Htwo$ (Fig. \ref{fig:one}a). Aside from its intrinsic interest, this geometric triality offers a powerful calculation tool. Whether embedded in $\Rthree$ or $\Hthree$, minimal surfaces are solutions of second order partial differential equations in $\bm{r}$ \cite{SI}, whereas {\em Elastica} curves embedded in space forms are fully integrable \cite{Langer1984}. With our dictionary in hand, computing the shape of lipid necks belongs now in the realm of possibility.

\begin{figure*}[t]
\includegraphics[width=0.75\textwidth]{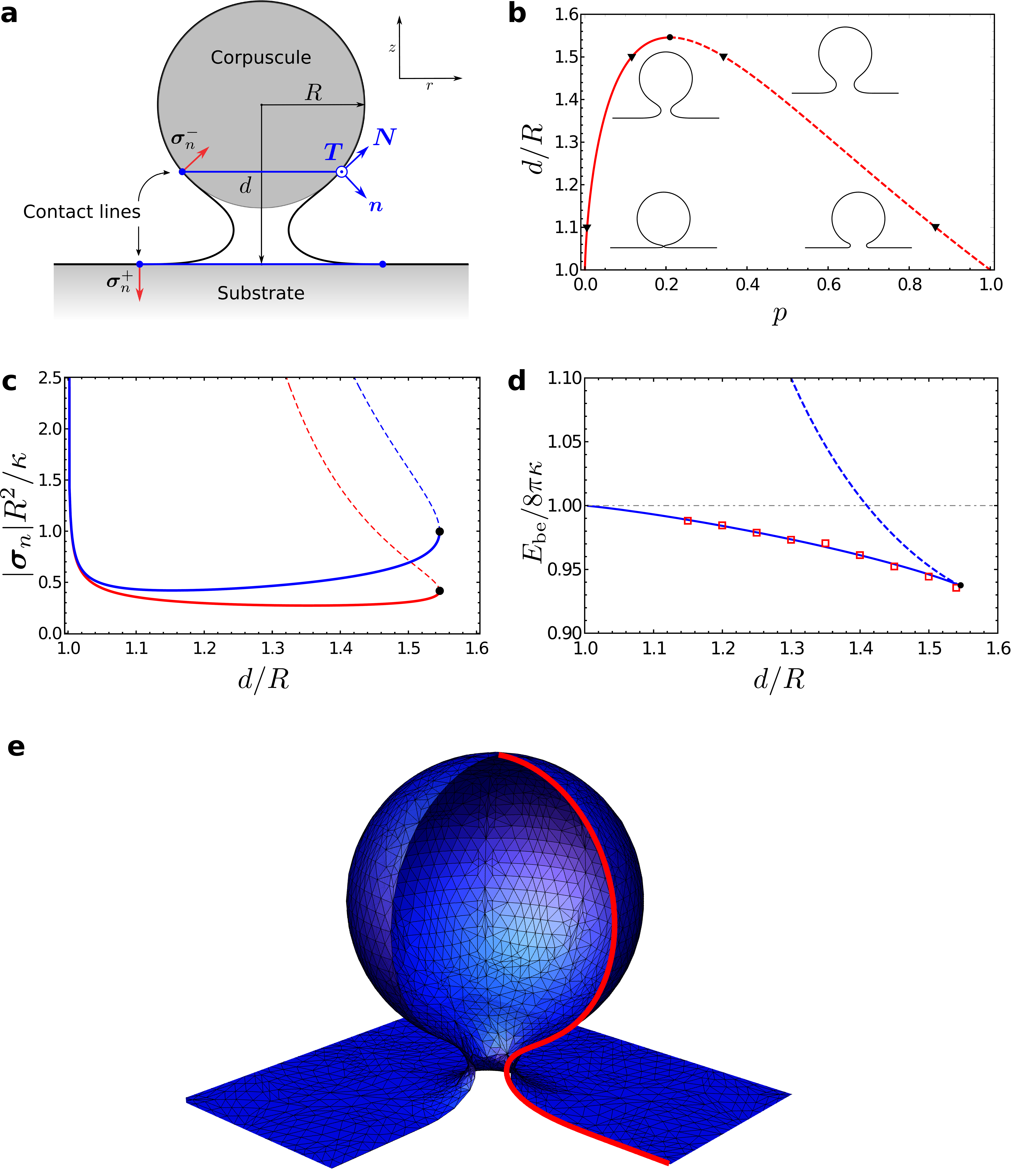}
\caption{ 
\textbf{a)} Variable definitions used in the main text to model the partial wrapping (configuration P in Fig. \ref{fig:one}b): a solid particle of radius $\Rpa$ is connected to a planar substrate via a neck. \textbf{b)} Sphere-plane distance, as given by \eqref{eq:sol distance}, of the P state as a function of the parameter $p$. For $\dpa<\dmax$ there are two $p$ values associated with the same $\dpa$ value. The solid/dashed portion of the line refers to stable/unstable solutions. Four insets show the membrane profile at the $p$ values highlighted by black triangles. The maximal distance $\dmax\simeq 1.546\Rpa$ is shown as a black dot. \textbf{c)} Normal stress at the contact lines as a function of $\dpa$. The blue (red) line refers to $\bm{\sigma}_n^-$ ($\bm{\sigma}_n^+$). The black dots show the values at $\dmax$. Dashed lines indicate the unstable branch. \textbf{d)} Bending energy as a function of distance. The partially wrapped state energy $\Epartial$ has two branches, one metastable (solid blue line) and one unstable (dashed blue line) which join at $\dmax$ (black dot). The red squares are values obtained numerically with Surface Evolver. The gray dot-dashed line shows the energy of the fully wrapped state $\Efull$. \textbf{e)} Comparison between the surface found by numerical minimization of \eqref{eq:E_be} (triangulated blue surface) and the analytical solution \eqref{eq:solution} (red line). See also Methods.
}
\label{fig:two}
\end{figure*}

\begin{figure*}[t]
\includegraphics[width=.75\linewidth]{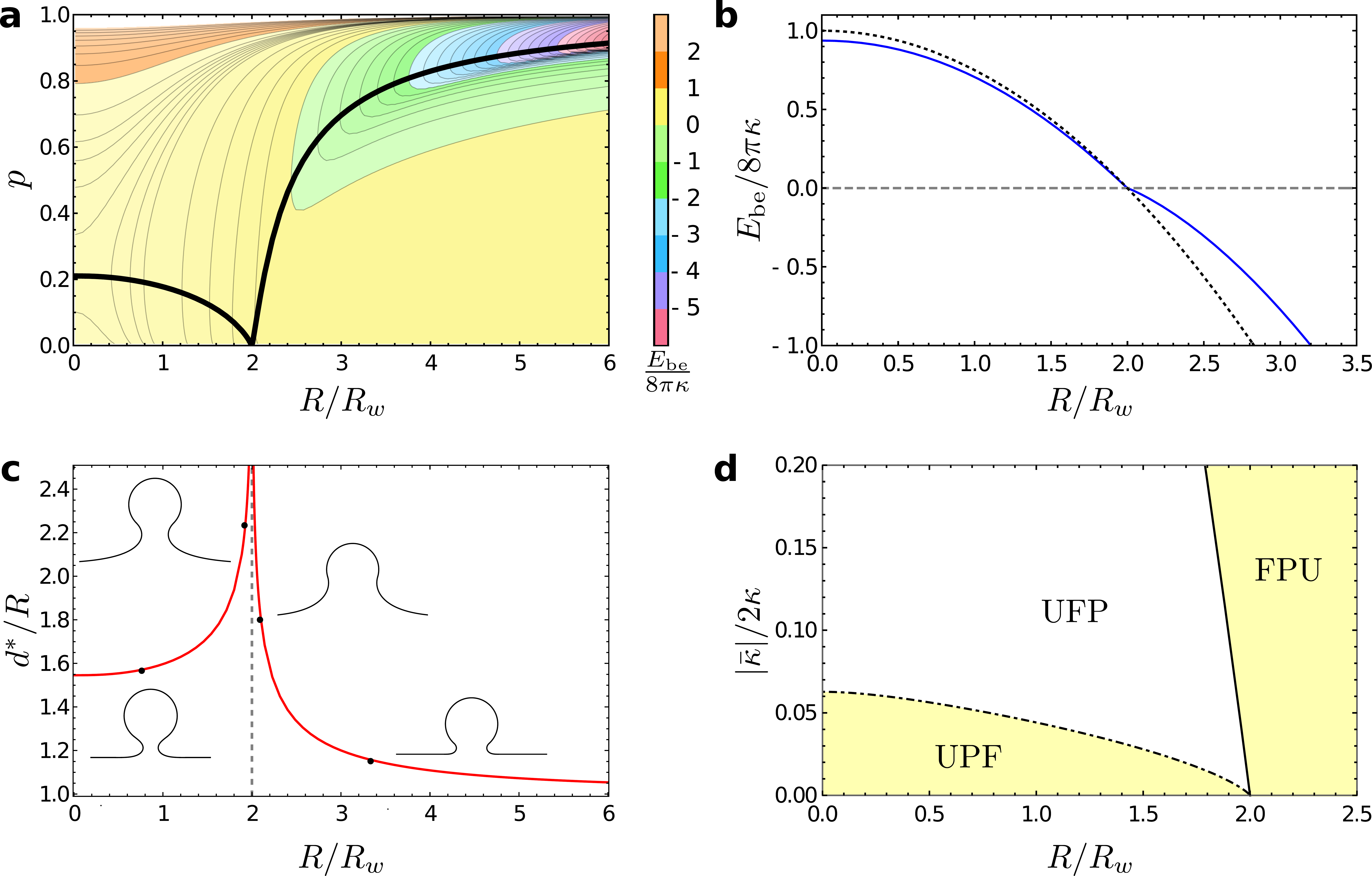}
\caption{ 
Effect of adhesion.
\textbf{a)} Energy landscape as a function of the parameter $p$ (which determines both $\bm{\sigma}_n^\pm$ and $\dpa$) and the adhesion length $\Rw$ of the partiall wrapped state. Different values of $\Epartial$ are color coded as shown in the bar legend. The solid dark line shows the location of the minimum for fixed $\Rw$.
\textbf{b)} Minimal energies as function of adhesion strength for the three different states shown in Fig. \ref{fig:one}b: unwrapped state U (dot-dashed gray), partially wrapped state P (solid blue) and fully wrapped F (dotted line). Note that $\Epartial$ has a discontinuity at $\Rw = 2\Rpa$. The optimal P state has intermediate energy values for all $w$. 
\textbf{c)} The red curve shows the equilibrium distance $\dmax$ as a function of $\Rw$ for P states. The divergence at $\Rw=\Rpa/2$ is due to the neck approaching a catenoid. The insets show the actual membrane shape attained for the values indicated by the black circles. 
\textbf{d)} Stability phase diagram of the three different states for different values of $w$ and $\bar{\kappa}$. The three letters are ordered from lowest to highest values of the energy, so e.g. UPF means that $\Eunwrap < \Epartial <\Efull$. Yellow-shaded regions are where metastable partial wrapping is an intermediate step between U and P.
}
\label{fig:three}
\end{figure*}

As an application of this approach, we consider a minimal model for necks formation during endo/exo-cytic processes, consisting of a spherical corpuscule of radius $\Rpa$ engulfed within an otherwise flat membrane spanning a two-dimensional plane (see Fig. \ref{fig:one}b). Adhesion between the corpuscule and the membrane will be introduced later, so that our task initially reduces to finding a Willmore {\em cylinder} bridging the gap between a plane and a sphere. The boundary conditions of this variational problem are determined by the balance of the tangential stress, $\bm{\sigma}_{t} = \kappa/2\,(C_{\parallel}^{2} - C_{\perp}^{2})\bm{N}$, and the bending moment, $\bm{m} = -2 \kappa H \bm{T}$, at the neck/sphere and neck/plane contact lines, where $\bm{T}$, $\bm{N}$ and $C_{\parallel}$ $(C_{\perp})$ are respectively the tangent and tangent-normal Darboux vectors (Fig. \ref{fig:two}a) and the parallel (transverse) normal curvature at the contact lines \cite{Capovilla2002}. Since there are no tangential forces, other than those caused by the membrane, {\em in-plane} stress balance requires that both principal curvatures vary continuously across the contact line. This rules out minimal surfaces in $\Rthree$, where $H=0$ everywhere, as possible configurations of the neck, despite them being obvious minimizers of the bending energy. Conversely, the normal stress $\bm{\sigma}_n = 2 \kappa \nabla_{\bm{N}} H \bm{n}$, with $\bm{n}$ the outward-pointing surface normal and $\nabla_{\bm{N}}=\bm{N}\cdot\nabla$, can be counterbalanced by {\em reaction forces} from the solid surfaces (red arrows in Fig. \ref{fig:two}a), thus enabling $\bm{\sigma}_{n}$ to vary discontinuously across both contact lines. Minimizing the right-hand side of either \eqref{eq:weyl bending} or \eqref{eq:elastica bending} yields the following explicit parametrization for the neck in cylindrical coordinates $r=(x^{2}+y^{2})^{1/2}$ and $z$:
\be
\begin{bmatrix}
r \\[5pt]
z
\end{bmatrix}
=
\frac{ \Rpa  \sinh 2 b_p(0)}{\cos \varphi  + a_p(\varphi) \cosh b_p(\varphi) }
\begin{bmatrix}
\sqrt{\frac{p}{1-p}}  \\[5pt]
 a_p(\varphi) \sinh b_p(\varphi)
\end{bmatrix}\,,
\label{eq:solution}
\ee
where 
\begin{subequations}
\begin{align}
a_p(\varphi) &= \sqrt{\frac{1}{1-p}-\sin^2\varphi} \;, \\
b_p(\varphi) &= \sqrt{\frac{p}{1-p}}\int_{\varphi}^{\pi/2} \D x\,\frac{1 - \frac{p}{1-p} \frac{1}{a_p(x)^2}}{\sqrt{p+\cos^2 x}}  \;.
\end{align}
\end{subequations}
The number $p\in[0,1]$ is a free parameter related with the residual normal stress $\bm{\sigma}_{n}$ and ultimately dictates the extension of the neck, whereas $\varphi\in[-\pi/2,\pi/2]$ depends on the hyperbolic arc-length $\tau$ appearing in \eqref{eq:elastica bending}: i.e. $\ell\, \D\varphi=\D\tau\,[(1-p)/(p+\cos^{2}\varphi)]^{-1/2}$, with $\varphi=\pm \pi/2$ corresponding to either the plane (plus) or the sphere (minus) contact lines. \eqref{eq:solution} describes a segment of the wave-like hyperbolic {\em Elastica} (Fig. \ref{fig:one}a), first found by Langer and Singer \cite{Langer1984a} in the context of embedded curves, and latter rediscovered in various areas of string theory, such as holographic duals of Wilson loops \cite{Gross1998,Drukker2006} and entanglement entropy \cite{Hirata2007,Krtous2014,Fonda2015a}, but never framed within the geometric triality depicted in Fig. \ref{fig:one}a. 

From the parametric form \eqref{eq:solution} we can derive analytical expressions of all physically relevant quantities, such as the relation between $p$ and the sphere-plane distance $\dpa$:
\be
\dpa = \Rpa \cosh 2 b_p(0) \;. 
\label{eq:sol distance}
\ee 
The ratio $\dpa/\Rpa$ is displayed in Fig. \ref{fig:two}b as a function of $p$. From this we see that $\dpa$ has a maximum $\dmax\simeq 1.546\Rpa$ at $p=p^*\simeq 0.210$, thus the neck can bridge the gap between the spherical and flat portions of the membrane only if these are sufficiently close to each other. For $\dpa>\dmax$, \eqref{eq:sol distance} has no real-valued solutions and the particle can only be {\em fully wrapped} by the membrane and disconnected from the plane (F state in Fig. \ref{fig:one}b), so that $\Ebe=\Efull = 8\pi\kappa$. By contrast, for $\Rpa<\dpa<\dmax$, \eqref{eq:sol distance} has two solutions (insets of Fig. \ref{fig:two}b), corresponding to different {\em partially wrapped} morphologies, whose bending energy, $E_{\rm be}=\Epartial = 8\pi\kappa e(p)$ with  
\be
e(p) = \frac{1}{\sqrt{1-p}} 
\int_0^{\pi/2}
\D x\, 
\frac{\cos^2 x}{\sqrt{p+\cos^2 x}}
\,,
\label{eq:sol energy}
\ee
is plotted in Fig. \ref{fig:two}c versus $\dpa/\Rpa$. Remarkably, the lower branch in Fig. \ref{fig:two}c has energy $\Epartial<\Efull$, indicating that, for any $\dpa<\dmax$, a partially wrapped sphere connected to the plane by a neck has lower bending energy than a fully wrapped sphere disconnected from the plane. Now, as the partially and fully wrapped configurations have different topologies, one must take into account possible energy costs associated with topological changes. Following Helfrich, these can be expressed as $\Eto=\bar{\kappa}\int {\D A\,K}=2\pi\bar{\kappa}\chi$, with $-2\kappa<\bar{\kappa}<0$ the Gaussian splay modulus \cite{Helfrich1973} and $K=C_{1}C_{2}$ the Gaussian curvature of $S$.  The energy difference between the partially and fully wrapped configurations is thus $\Delta E = (\Epartial-\Efull)-4\pi\bar{\kappa}$, where we have used the fact that $\chi=2$ for a sphere. As $\Epartial<\Efull$ for $\Rpa<\dpa<\dmax$, the partially wrapped configuration is of lower energy for $|\bar{\kappa}|/2\kappa<1-e(p^{*})=0.063$, or {\em metastable} with respect to full wrapping otherwise. 

To test the significance of our predictions, we have complemented these results with numerical simulations. These are performed via the Surface Evolver (SE) \cite{Brakke1992} (Methods and Fig. \ref{fig:two}e), without making any assumption about axial symmetry nor on the location of the contact lines. The numerically found bending energy is marked by red squares in Fig. \ref{fig:two}d and is in excellent agreement with our analytical result, thus confirming that necks are locally stable minima of the bending energy \eqref{eq:E_be}.  Furthermore, as we stressed earlier, any configuration involving a partial wrapping on the sphere requires reaction forces to compensate the residual normal normal force $\bm{F}_{n}=\oint \D s\, \bm{\sigma}_{n}$ resulting from the discontinuity in the tangent-normal derivative of the mean curvature. Using \eqref{eq:solution}, this can be calculated to be: $\bm{F}_n^\pm = 2\pi \bm{\sigma}_n^\pm \Rpa \left(\varepsilon_\pm +\sinh^{-2} 2 b_p(0)\right)^{-1/2}$, where $\bm{\sigma}_{n}^{\pm}=2\kappa\sqrt{p}/[\Rpa^{2}(p-1)][\varepsilon_{\pm}+\sinh^{-2}2b_p(0)]\,\bm{n}$ and $\varepsilon_+=0$ and $\varepsilon_-=1$ corresponds to the neck-plane and neck-sphere contact line respectively. A plot of $|\bm{\sigma}_{n}^{\pm}|$ versus $\dpa/\Rpa$ is shown in Fig. \ref{fig:two}c. Taking $\kappa \approx 25\,k_{B}T$ \cite{Dimova2014} one finds that a nanoparticle of radius $\Rpa=500\,{\rm nm}$ would experience normal forces $|\bm{F}_n^+| \approx  1.2 \,{\rm pN}$ and $|\bm{F}_n^-| \approx 1.9\,{\rm pN}$ at $\dpa=\dmax$. 

Finally, we consider the practically relevant case in which the membrane-particle interaction is mediated by adhesive forces. The latter, can be accounted via the additional energy cost $\Ead = -w \Aad$, with $w>0$ the average adhesion strength and $\Aad$ the portion of the membrane in contact with the substrate \cite{Seifert1990}. Since this does not influence the energetics of the free portion of the membrane, adhesion does not break conformal invariance nor alter the structure of the solution as given in \eqref{eq:solution}, but does modify the boundary conditions at the contact lines by constraining the magnitude of the transverse curvature: i.e. $\ktheta=\Rpa^{-1}-\Rw^{-1}$, where $\Rw=\sqrt{\kappa/2w}$ \cite{Deserno2007}. This scenario is illustrated in Fig. \ref{fig:three}, where, for simplicity, we restrict ourselves to adhesive forces at the neck-sphere contact line \cite{SI}. Fig. \ref{fig:three}a shows the bending energy $\Ebe$ versus $p$ and the adhesion strength, expressed in terms of the dimensionless ratio $\Rpa/\Rw\sim\sqrt{w}$. The solid curve marks the locus of the energy minima for fixed $\Rw$ values (so that the leftmost point of the curve lies at $p=p^*$, see Fig. \ref{fig:two}b and \ref{fig:two}d). Fig. \ref{fig:three}b shows a comparison between the bending energies of the partially (blue) and fully wrapped (dashed) configurations versus $\Rpa/\Rw$. Being characterized by the largest possible contact area, the fully wrapped configuration is energetically favored in the presence of large adhesive forces, but this trend is reversed for $\Rpa<2\Rw$, when bending overweight other energy costs. Exactly at the crossover point, the optimal sphere-plane distance $\dmax$ diverges (Fig. \ref{fig:three}c), as the free portion of the membrane approaches a catenoid. This singular behaviour is reflected in a discontinuity of $\Epartial$, yet for all $w$ values the P state has an intermediate energy between the U and P configurations. Combining these observations with the earlier discussion on topological contributions, we obtain the phase diagram shown in Fig. \ref{fig:three}d. For small Gaussian splay and large adhesion strength values, the partially wrapped configuration has intermediate energy between U and F. Since there is no locally stable minima of the energy other than U, P and F and being partial wrapping an inevitable intermediate step in any process involving the penetration of the cell membrane by an external or internal corpuscule, we infer that the metastability of lipid necks will certainly play a significant role in the dynamics of endo/exo-cytosis.

In conclusion, we have demonstrated that finite-size lipid necks, namely highly curved features of lipid membranes often functioning as bridges between other compartments, are metastable and can exist for a finite, but potentially long time, even in the absence of stabilizing mechanisms. Our approach is rooted on a geometric triality, linking Willmore surfaces in $\Rthree$, minimal surfaces in $\Hthree$ and the Euler {\em Elastica} in $\Htwo$, and capitalizes on concepts found in disconnected areas of mathematics and high energy physics, but never united within a common framework, nor exploited in biophysics. 
This framework allows us to analytically describe the geometrical and physical properties of a spherical corpuscule adhering onto a flat membrane and discover that, in addition of being metastable, partially wrapped configurations are in fact energetically intermediate between unwrapped and fully wrapped configurations for a wide range of material parameter values (see Fig. \ref{fig:three}d) and thus are expected to strongly influence any dynamical process involving the adsorption/emission of particle by lipid membranes. Our analysis includes considerations on out-of-plane stresses due to curvature discontinuities across contact lines, finding that sub-mircometer particles require forces in the piconewton range to sustain partial wrapping. In both eukaryotes and prokaryotes, these forces can be generated by protein scaffolds, such as cytoskeletal filaments and motors. Furthermore, our findings finally resolve a dilemma in the existing theoretical literature, where partial wrapping had been observed in molecular dynamics simulations of tensionless lipid bilayers \cite{Ruiz-Herrero2012,Spangler2016}, but never recovered in the framework of Helfrich's theory.
We also note that in the recent experimental work of Ref. \cite{Spanke2020} a biomimetic system matching our model assumptions was studied, were nanoparticle were pushed through a symmetric bilayer with an optical tweezer: this is the ideal setting to study metastable states as the ones described here. Whether to understand how organelles generate liposomes, how viruses can enter host cells or how nanoparticles can deliver drugs, a fundamental comprehension of membrane mechanics remains essential. Our work, exemplified in the triality diagram of Fig. \ref{fig:one}b, offers a new perspective to look and the complex mechanics of lipid membranes.

\section{Methods}
\subsection{Details on the numerical simulation}

To reproduce the stable branch of the energy shown in Fig. \ref{fig:two}d, we implemented a 3D simulation of the system shown in Fig. \ref{fig:two}a with Surface Evolver (SE) \cite{Brakke1992}. The initial surface was constructed as the union of a prism and an polygonal annulus, while the constraints of the substrate and the particle were implement as hard implicit level-set inequalities. Through mesh refinements and gradient-descent methods (derived from the vertex-wise derivative of a discretize version of \eqref{eq:E_be}), the triangulated surface was made to flow towards locally stable solutions as shown e.g. in Fig. \ref{fig:two}e. Due the high-derivative nature of the Willmore energy and of the Willmore flow, convergence is relatively slow. Code available at \href{https://github.com/pierfonda/evolverneck}{github.com/pierfonda/evolverneck.}

\bibliography{references}{}

\end{document}